\documentclass[prb,twocolumn,superscriptaddress,a4paper,twoside]{revtex4}

\usepackage{amsmath}
\usepackage{amssymb}
\usepackage{todonotes}
\usepackage{nicefrac}
\usepackage{graphics}
\usepackage{color}
\usepackage{graphicx}
\usepackage{verbatim}
\usepackage{float}

\makeatletter
\newcommand*{\rom}
[1]{\expandafter\@slowromancap\romannumeral #1@}
\makeatother

\begin{document}

\title{Quasiparticle wavefunction and its equation of motion}

\author{
F. Aryasetiawan}
\affiliation{
Department of Physics, Division of Mathematical Physics, 
Lund University, Professorsgatan 1, 223 63, Lund, Sweden}
\affiliation{LINXS Institute of advanced Neutron and X-ray Science (LINXS),
IDEON Building: Delta 5, Scheelevägen 19, 223 70 Lund, Sweden}
\author{
K. Karlsson}
\affiliation{Department of Engineering Sciences, 
University of Sk\"{o}vde, SE-541 28 Sk\"{o}vde, Sweden}

\begin{abstract}
The quasiparticle wavefunction of a many-electron system is traditionally
defined as the eigenfunction of the quasiparticle eigenvalue equation involving
the self-energy. In this article a new concept of a quasiparticle wavefunction
is derived from the general definition of the Green function without
reference to self-energy. The proposed quasiparticle wavefunction
can decay in time, and
in contrast to the traditional one it
contains not only the main quasiparticle mode but also other modes
due to coupling to collective excitations in the system.
In the recently developed dynamical exchange-correlation potential
formalism, the new definition of a quasiparticle wavefunction leads
to an equation of motion with an effective field, which appears to have
a simple form expected to be amenable to realistic approximations. 
A simple model for the effective potential is proposed, which is 
suitable for electron-gas-like materials such as the alkali.
\end{abstract}

\maketitle

\section{Introduction}

Angle-resolved photoemission spectra of a large number of materials
exhibit a generic feature characterized by the presence of a main peak
close to the chemical potential and additional incoherent features
at higher binding energies. The main peak is usually referred to as
the quasiparticle peak, whereas the incoherent features can be traced back
to the coupling of electrons to collective excitations, typically plasmons.
In magnetic systems the electrons can be coupled to spin excitations
or magnons giving rise
to features at low binding energies such as kinks in the band dispersion.

The concept of a quasiparticle was first introduced by Landau in the mid 1950's
in his famous phenomenological Fermi liquid theory \cite{fetter-walecka}. Landau quasiparticles are
restricted to those long-lived excitations at the Fermi level, but later
development of Green function theory offers a more general concept of a
quasiparticle. In metals a quasiparticle at the Fermi level does indeed have
an infinite lifetime as predicted by Landau, whereas away from the Fermi level
the quasiparticle usually acquires a finite lifetime, so that it
decays with time.
In the language of self-energy, the lifetime is inversely
proportional to the imaginary part of the self-energy. 

Traditionally, the quasiparticle is associated with a one-particle
wavefunction defined to be an eigenfunction of
the quasiparticle equation \cite{lundqvist}
in which the self-energy together with the Hartree and external potentials
form the total potential of the Hamiltonian:
\begin{align}\label{eq:QPtrad}
   & \left[-\frac{1}{2}\nabla^2 + V_\mathrm{H}(r) +V_\mathrm{ext}(r)
    \right] \Psi_k(r,E_k)
    \nonumber\\
    &+ \int dr' \Sigma(r,r';E_k) \Psi_k(r',E_k)
    = E_k \Psi_k(r,E_k).
\end{align}
Here, the notation $r=(\mathbf{r},\sigma)$ denoting both space and
spin variables is used. 
Atomic units (a.u.) are used throughout the article except indicated
otherwise.
$V_\mathrm{H}(r)$ is the Hartree potential, $V_\mathrm{ext}(r)$ is
the external potential, and $\Sigma(r,r';E_k)$ is the self-energy
calculated at the quasiparticle energy $E_k$. The quasiparticle
wavefunction is given by $\Psi_k(r,E_k)$ which is normalized to unity.
Since the Hamiltonian is not Hermitian, quasiparticles with different
energies are not in general orthonormal.

In this article, a different definition of the quasiparticle wavefunction
is proposed, which arises naturally from the definition of the Green function.
In contrast to the traditional one defined in Eq. (\ref{eq:QPtrad}),
the proposed quasiparticle wavefunction decays in time and contains
excitation modes associated with the addition or removal of an electron
from a many-electron system. It contains not only
the quasiparticle mode but also other modes due to the coupling
with collective excitations such as plasmons. 
In the recently developed dynamical
exchange-correlation (xc) potential 
formalism \cite{aryasetiawan2022a,aryasetiawan2022b,karlsson2023,
aryasetiawan2023,zhao2023} the quasiparticle wavefunction
fulfills a one-particle time-dependent equation of motion.
From its solutions, the Green function can then be constructed.

\section{Quasiparticle wavefunction}

The time-ordered zero-temperature Green function for a system
in equilibrium is defined as \cite{fetter-walecka,negele}
\begin{align}
    iG(r,r';t)=\langle\Psi_0| \mathrm{T}\hat{\psi}(rt) 
    \hat{\psi}^\dagger(r'0)|\Psi_0\rangle,
\end{align}
in which $\mathrm{T}$ is the time-ordering symbol, $|\Psi_0\rangle$
is the ground state, and $\hat{\psi}(rt)$ is the field operator in
the Heisenberg picture,
\begin{align}
    \hat{\psi}(rt)=e^{i\hat{H}t} \hat{\psi}(r) e^{-i\hat{H}t}.
\end{align}

Choosing a complete set of one-particle orbitals $\{\varphi_k\}$
the field operator can be expanded as
\begin{align}
    \hat{\psi}(r)=\sum_k \varphi_k(r)\hat{c}_k,
    \qquad \hat{\psi}^\dagger(r)=\sum_k \hat{c}^\dagger_k\varphi^*_k(r),
\end{align}
where $\hat{c}_k$ and $\hat{c}^\dagger_k$ are the annihilation
and creation operators associated with orbital $\varphi_k$.
The Green function can then be written as
\begin{align}\label{eq:Gorb}
    G(r,r';t) &= \sum_{kk'} \varphi_{k}(r)G_{kk'}(t)\varphi^*_{k'}(r'),
\end{align}
where
\begin{align}
    iG_{kk'}(t) =\langle\Psi_0| \mathrm{T}\hat{c}_k(t) 
    \hat{c}^\dagger_{k'}(0)|\Psi_0\rangle.
\end{align}

Defining \cite{aryasetiawan2023}
\begin{align}\label{eq:QPwf}
    \psi^*_k(r',t)&=\sum_{k'} G_{kk'}(t)\varphi^*_{k'}(r'),
\end{align}
the Green function can be expressed as
\begin{align}\label{eq:GQP}
    G(r,r';t) &=  \sum_k \varphi_k(r) \psi^*_k(r',t).
\end{align}
It also follows that
\begin{align}
    \psi^*_k(r',t)
    &=\int dr\, \varphi^*_k(r) G(r,r';t),
\end{align}
\begin{align}
    G_{kk'}(t)=\int dr'\,\psi^*_k(r',t) \varphi_{k'}(r').
\end{align}
Thus, $\psi^*_k(r',t)$ for all $k$ contain all the necessary information
needed to reconstruct the Green function.
It is argued that $\psi^*_k(r',t)$
can be interpreted as a quasiparticle wavefunction.

For a noninteracting system, $G_{kk'}(t)=G^0_{k}(t)\delta_{kk'}$ so that
\begin{align}
    G^0(r,r';t) &= \sum_{k} \varphi_{k}(r)G^0_k(t)\varphi^*_{k}(r'),
\end{align}
in which
\begin{align}
    G^0_k(t) = i \theta(-t) \theta(\mu-\varepsilon_k)
                   e^{-i\varepsilon_k t} 
            - i\theta(t) \theta(\varepsilon_k-\mu)
                   e^{-i\varepsilon_k t}.
\end{align}
Here, $\varepsilon_k$ is the orbital energy 
and $\mu$ is the chemical potential.
The noninteracting quasiparticle wavefunction is then
\begin{align}
    \psi^{0*}_k(r,t)=G^0_k(t)\varphi^*_{k}(r).
\end{align}
It is quite evident that
\begin{align}
    |G^0_k(t)|^2=1,
\end{align}
which implies that for a noninteracting system in equilibrium, 
the quasiparticle does not decay, as expected.

%
%
For interacting systems, 
$G_{kk'}(t)$ can be interpreted as the expansion coefficient
of $\psi^{0*}_k(r,t)$ in the base orbital $\varphi_{k'}$. To see this,
consider the case $t<0$:
\begin{align}
    |G_{kk}(t)|^2 &= \left|
    \langle \Psi_0 | \hat{c}_k^\dagger \hat{c}_k(t) | \Psi_0\rangle \right|^2
    \nonumber\\
    &= \left|
    \langle \Psi_0 | \hat{c}_k^\dagger e^{i\hat{H}t}\hat{c}_k | \Psi_0\rangle
    \right|^2.
\end{align}
Let $|\phi_k \rangle = \hat{c}_k | \Psi_0\rangle$. 
For fermions, due to the Pauli principle, 
the expectation value of the number operator
$\hat{n}_k=\hat{c}^\dagger_k\hat{c}_k$
cannot exceed unity so that
\begin{align}
    \langle \phi_k |\phi_k\rangle 
    = \langle \Psi_0 |\hat{n}_k |\Psi_0\rangle \leq 1.
\end{align}
Since the operator $e^{i\hat{H}t}$ is unitary, it follows that
\begin{align}
    |G_{kk}(t)|^2 = \left|
    \langle \phi_k |e^{i\hat{H}t} | \phi_k\rangle
    \right|^2 \leq 1.
\end{align}
The operator $e^{i\hat{H}t}$ rotates the state $ | \phi_k\rangle$ so that
the magnitude of the overlap of the rotated state with
$\langle \phi_k|$ must be less than $\langle \phi_k |\phi_k\rangle$.

For a given $t$, let $S$ be the unitary matrix that diagonalizes $G$:
%
%
%
\begin{align}
    G_{kk'}(t) = \sum_{k_1} S_{kk_1} \widetilde{G}_{k_1}(t) S^\dagger_{k_1k'}.
\end{align}
One finds
\begin{align}
   \sum_{k'} |G_{kk'}(t)|^2 &= \sum_{k'} \sum_{k_1k_2}
   S_{kk_1} \widetilde{G}_{k_1}(t) S^\dagger_{k_1k'} S_{k'k_2}
   \widetilde{G}^*_{k_2}(t) S^\dagger_{k_2k}
   \nonumber\\
   &=\sum_{k_1}S_{kk_1} |\widetilde{G}_{k_1}(t)|^2 S^\dagger_{k_1k}
   \nonumber\\
   &=\sum_{k_1}|S_{kk_1}|^2 |\widetilde{G}_{k_1}(t)|^2.
\end{align}
Since $\sum_{k_1}|S_{kk_1}|^2=1$
and $|\widetilde{G}_{k_1}(t)|^2 \leq 1$ it follows that
\begin{align}
    \sum_{k'} |G_{kk'}(t)|^2 \leq 1.
\end{align}
This result shows that
\begin{align}\label{eq:intpsik2}
    \int dr\, |\psi_k(r,t)|^2 = \sum_{k'} |G_{kk'}(t)|^2 \leq 1
\end{align}
and supports the interpretation of $\psi^*_k(r,t)$ in Eq. (\ref{eq:QPwf})
as quasiparticle wavefunction. $G_{kk'}(t)$ can therefore be understood as the
expansion coefficient of the quasiparticle wavefunction
$\psi^*_k(r,t)$ in the base orbital
$\varphi^*_{k'}$. A similar analysis can be performed for $t>0$.
Extension to a finite temperature amounts to replacing
the ground-state expectation value by a thermal average, and
extension to a nonequilibrium case can be found in Appendix \ref{app:noneq}.
The property in Eq. (\ref{eq:intpsik2}) may be seen from a different
point of view \cite{dvorak2021}.

This definition of the quasiparticle
wavefunction is general, unrelated to any formalism for calculating the
Green function, be it the traditional self-energy or
the dynamical xc potential formalism described later.
Unlike the traditional quasiparticle wavefunction in
Eq. (\ref{eq:QPtrad}) defined as the eigenfunction corresponding
to the quasiparticle energy, here the quasiparticle wavefunction is
time dependent and contains not only the quasiparticle mode but also other
modes arising from the coupling to collective excitations in the system such 
as plasmons. These modes appear as incoherent features in angle-resolved
photoemission spectra.


\section{Quasiparticle equation in the dynamical xc
potential formalism}

In the dynamical xc potential formalism, the
equation of motion of the Green function is given by \cite{aryasetiawan2022a}
\begin{align}
    \left[ i\partial_t -h(r)-V_\mathrm{xc}(r,r';t)\right]G(r,r';t)=
    \delta(r-r')\delta(t),
\end{align}
in which
\begin{align}
    h(r)=-\frac{1}{2}\nabla^2 + V_\mathrm{H}(r) +V_\mathrm{ext}(r).
\end{align}
$V_\mathrm{xc}$ is the dynamical xc potential
which is the Coulomb potential of the xc hole
$\rho_\mathrm{xc}$,
\begin{align}
    V_\mathrm{xc}(r,r';t)
    =\int dr'' v(r-r'')\rho_\mathrm{xc}(r,r',r'';t).
\end{align}
In the limit $r'=r$ and $t=0^-$, the exchange part of $\rho_\mathrm{xc}$
reduces to the Slater exchange hole \cite{slater1951,slater1968},
\begin{align}
    \rho_\mathrm{x}(r,r,r'';0^-)=\rho_\mathrm{x}^\mathrm{Slater}(r,r''),
\end{align}
provided a noninteracting Green function is used
when calculating $\rho_\mathrm{x}$.
$V_\mathrm{xc}$ can then be viewed as a generalization of the
Slater exchange potential, which includes the dynamical effects of
exchange and correlations. As in the case of
the Slater exchange hole, the dynamical
xc hole also fulfills a sum rule:
\begin{align}
    \int dr'' \rho_\mathrm{xc}(r,r',r'';t) 
    = -\theta(-t)\delta_{\sigma\sigma''}.
\end{align}

It is often useful to work with an orbital basis as in
Eq. (\ref{eq:Gorb}). With an orbital basis, the equation
of motion takes the form
\begin{equation}
i\frac{\partial}{\partial t}G_{ij}(t)-\sum_{k}h_{ik}G_{kj}(t)-\sum
_{kl}V_{ik,lj}^{xc}(t)G_{kl}(t)=\delta_{ij}\delta(t),
\end{equation}
where $G_{ij}$ and $h_{ik}$ are the matrix elements of $G$ and $h$ in the
orbitals and
\begin{equation}\label{eq:Vxc_ijkl}
V_{ik,lj}^\mathrm{xc}(t)
=\int d^{3}rd^{3}r^{\prime}\text{ }
\varphi_{i}^{\ast}(r)\varphi_{k}(r)V_\mathrm{xc}(r,r^{\prime};t)
\varphi_{l}^{\ast}(r^{\prime})\varphi_{j}(r^{\prime}).
\end{equation}
%

Writing
\begin{equation}
    \Delta V(r,r';t) = V_\mathrm{xc}(r,r';t) - V_\mathrm{xc}^\mathrm{0}(r),
\end{equation}
where $V_\mathrm{xc}^\mathrm{0}$ 
is a mean-field xc potential,
which could be chosen to be the Kohn-Sham xc potential
\cite{kohn1965,jones1989,becke2014}.
Choosing the orbitals to be those of the mean-field ones,
one finds from the
equation of motion with $t\neq 0$
\begin{align}\label{eq:EOMQP}
    \sum_{k} \left[ i\partial_t -\varepsilon_k- \Delta V(r,r';t)\right]
     \varphi_k(r) \psi^*_k(r',t) =0.
\end{align}
$\varepsilon_k$ is the eigenenergy of the mean-field orbital $\varphi_k$.
Multiplying on the left by $\varphi^*_q(r)$ and integrating over $r$ yields
(renaming $r'$ by $r$ after integration),
\begin{align}
    (i\partial_t-\varepsilon_q)\psi^*_q(r,t) 
    -\sum_{k} \Delta V_{qk}(r,t) \psi^*_k(r,t)=0,
\end{align}
in which
\begin{align}\label{eq:DeltaV}
    \Delta V_{qk}(r,t) = \int dr'\, \varphi^*_q(r')\varphi_k(r')\Delta V(r',r;t).
\end{align}
The last term with $k\neq q$,
\begin{align}
\sum_{k\neq q} \Delta V_{qk}(r,t) \psi^*_k(r,t),
\end{align}
can be interpreted as the coupling with other quasiparticles with
quantum number different from $q$.

The quasiparticle equation of motion can be recast as
\begin{align}\label{eq:EOMXi}
    \left[i\partial_t-\varepsilon_q-\Xi_q(r,t)\right]\psi^*_q(r,t) =0,
\end{align}
where
\begin{align}\label{eq:Xi}
    \Xi_q(r,t) =\frac{1}{\psi^*_q(r,t)}
    \sum_{k} \Delta V_{qk}(r,t) \psi^*_k(r,t)
\end{align}
is an effective $q$-dependent potential, which corrects the mean-field
eigenvalue and incorporates dynamical xc effects beyond the
mean-field description. Note that
each quasiparticle wavefunction
sees its own potential. 
The formal solution is given by
\begin{align}
    \psi^*_q(r,t)=\psi^*_q(r,0)
    e^{-i\varepsilon_q t 
    -i\int_0^t dt'  \Xi_{q}(r,t')}.
\end{align}
%
Once solved for all $q$, 
the Green function can be constructed from Eq. (\ref{eq:GQP}).
When the dynamical effects of exchange and correlations encapsulated
in $\Xi$ are switched off, the quasiparticle wavefunction
returns to the noninteracting reference orbital.
The decay of the quasiparticle wavefunction is consistent with
the fact that the dynamical xc potential is
not Hermitian in general. 
The imaginary part of $\Xi_q(r,t)$ causes the quasiparticle
wavefunction to decay with time in an interacting many-electron system.

The quasiparticle equation of motion depends on
a potential $\Xi_q(r,t)$ which in turn
depends on the quasiparticle states. 
To simplify the equation of motion, a possible
approximation is to take the quasiparticle average of
$\Xi_q$ over all $q$:
\begin{align}
    \Xi^\mathrm{eff}(r,t)&= \sum_q 
    \frac{|\psi_q(r,t|^2}{\rho_\mathrm{QP}(r,t)}
     \Xi_{q}(r,t)
     \nonumber\\
     &= \frac{1}{\rho_\mathrm{QP}(r,t)}
     \sum_{qk} \psi_q(r,t) \Delta V_{qk}(r,t)
     \psi^*_k(r,t),
\end{align}
where
\begin{align}
    \rho_\mathrm{QP}(r,t)=\sum_q |\psi_q(r,t)|^2.
\end{align}
This approximation is inspired by Slater's orbital average
of the local exchange potential. The equation of motion
now becomes
\begin{align}
    \left[i\partial_t-\varepsilon_q-
    \Xi^\mathrm{eff}(r,t)\right]\psi^*_q(r,t) =0,
\end{align}
in which the effective potential is $q$-independent.
However, for a homogeneous system such as the electron gas, 
the $q$-dependence is important, without which
the quasiparticle dispersion will remain unchanged.
A further simplification can be made by assuming that the quasiparticle
states appearing
in $\Xi^\mathrm{eff}(r,t)$ are given by the noninteracting quasiparticle states.

\section{Examples}

Two examples will be considered to illustrate the concept of
quasiparticle wavefunction proposed in the present article.
The two examples are the Hubbard dimer \cite{aryasetiawan2022a} 
and the homogeneous electron gas \cite{lundqvist,karlsson2023},
which represent two opposite extremes of many-electron systems, to illustrate
that the concept is quite general.

\subsection{Hubbard dimer}

Although it is very simple and analytically solvable, 
the half-filled Hubbard dimer
with total $S_z=0$ has the merit of having some of the essential
ingredients of correlated electrons. It also illustrates that
the proposed quasiparticle wavefunction
is not restricted to continuous systems but
can be carried over to finite systems and lattice models.

The Hamiltonian of the Hubbard dimer in standard
notation is given by
\begin{equation}
\hat{H}=-\Delta\sum_{i\neq j,\sigma}\hat{c}_{i\sigma}^{\dag}\hat{c}_{j\sigma}%
+U\sum_{i}\hat{n}_{i\uparrow}\hat{n}_{i\downarrow}.
\end{equation}
In some cases, it is useful to choose the bonding (B) and antibonding (A)
orbitals as the basis:
\begin{align}
    \phi_\mathrm{B}=\frac{1}{\sqrt{2}}\left( \varphi_1 +\varphi_2\right),
    \quad
    \phi_\mathrm{A}=\frac{1}{\sqrt{2}}\left( \varphi_1 -\varphi_2\right).
\end{align}
In this basis the Green function is diagonal and the
xc potential is given by
\begin{align}
     V^\mathrm{xc}_\mathrm{BB,BB}=V^\mathrm{xc}_\mathrm{AA,AA}=
    \frac{1}{2}(V^\mathrm{xc}_{11,11}+V^\mathrm{xc}_{12,21}),\\
    V^\mathrm{xc}_\mathrm{AB,BA}=V^\mathrm{xc}_\mathrm{BA,AB}=
    \frac{1}{2}(V^\mathrm{xc}_{11,11}-V^\mathrm{xc}_{12,21}).
\end{align}
The definition of the matrix elements is given in Eq. (\ref{eq:Vxc_ijkl}).
In the bonding-antibonding basis 
the equation of motion of the Green function becomes
\begin{align}
    \left[ i\partial_t -\varepsilon_\mathrm{B}- V^\mathrm{xc}_\mathrm{BB,BB}(t)
    \right] G_\mathrm{B}(t) -V^\mathrm{xc}_\mathrm{BA,AB}(t) G_\mathrm{A}(t)
    =\delta(t),
\end{align}
\begin{align}
    \left[ i\partial_t -\varepsilon_\mathrm{A}- V^\mathrm{xc}_\mathrm{AA,AA}(t)
    \right] G_\mathrm{A}(t) -V^\mathrm{xc}_\mathrm{AB,BA}(t) G_\mathrm{B}(t)
    =\delta(t),
\end{align}
where
\begin{align}
    \varepsilon_\mathrm{B}=-\Delta,\quad \varepsilon_\mathrm{A}=\Delta.
\end{align}
For the Hubbard dimer, the quasiparticle wavefunctions are simply
$G_\mathrm{B}\phi_\mathrm{B}$ and $G_\mathrm{A}\phi_\mathrm{A}$. They 
can be solved analytically yielding
\begin{align}
iG_\text{A}(t  &  >0)=a_{0}^2 (1+x)^{2}\exp(-i\varepsilon_{0}^{+}t),
\end{align}
\begin{align}
iG_\text{A}(t  &  <0)=-a_{0}^2 (1-x)^{2}\exp(i\varepsilon_{1}^{-}t),
\end{align}
\begin{align}
iG_\text{B}(t  &  >0)=a_{0}^2 (1-x)^{2}\exp(-i\varepsilon_{1}^{+}t),
\end{align}
\begin{align}
iG_\text{B}(t  &  <0)=-a_{0}^2 (1+x)^{2}\exp(i\varepsilon_{0}^{-}t),
\end{align}
where
\begin{equation}
    a_0^2 = \frac{1}{2(1+x^2)},\qquad x=-\frac{E_0}{2\Delta}.
\end{equation}
$E_0$ is the ground-state energy, 
\begin{align}
E_{0}  &  =\frac{1}{2}\left(  U-\sqrt{U^{2}+16\Delta^{2}}\right) .
\end{align}

The orbital energies are given by
\begin{align}
    \varepsilon_0^-=-\Delta-E_0,\quad \varepsilon_0^+=U-\Delta-E_0,\\
    \varepsilon_1^-=\Delta-E_0, \quad \varepsilon_1^+=U+\Delta-E_0.
\end{align}

For $t>0$, one obtains \cite{aryasetiawan2022b}
\begin{align}\label{eq:VxcBB}
     V^\mathrm{xc}_\mathrm{BB,BB}(t>0)=\frac{\alpha U}{2}
     \frac{1-\alpha^2 e^{-i4\Delta t}}{1-\alpha^4 e^{-i4\Delta t}},
\end{align}
\begin{align}\label{eq:VxcAB}
     V^\mathrm{xc}_\mathrm{AB,BA}(t>0)=\frac{\alpha U}{2}
     \frac{(1-\alpha^2) e^{-i2\Delta t}}{1-\alpha^4 e^{-i4\Delta t}},
\end{align}
where
\begin{equation}\label{alpha}
\alpha=\frac{1-x}{1+x}.
\end{equation}%
Due to particle-hole symmetry, $V^\mathrm{xc}(-t)=-V^\mathrm{xc}(t)$.
The parameter $\alpha$ provides a measure of correlations. The larger
$\alpha$, the stronger the correlation.

\begin{figure}[t]
\begin{center} 
\includegraphics[scale=0.6, viewport=3cm 8cm 17cm 20cm, clip]
{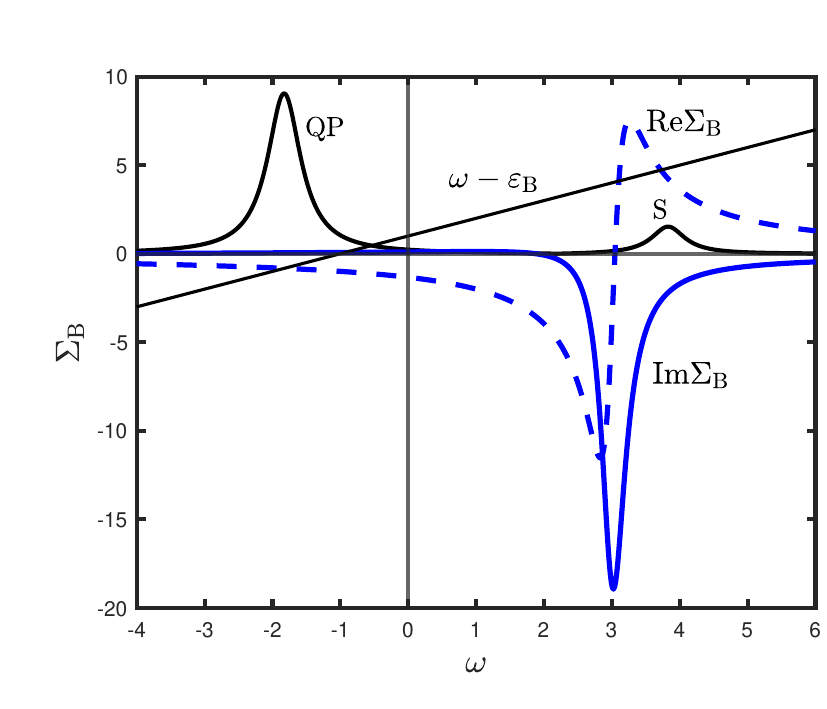}
\caption{The real and imaginary parts of the self-energy and the 
spectral function of the Hubbard dimer
for $U=4$ and $\Delta=1$ in the bonding orbital.
The energies at which the straight line $\omega-\varepsilon_\mathrm{B}$
crosses the real part of the self-energy Re$\Sigma_\mathrm{B}$
correspond to the main peak (QP) in the spectral function
and an additional structure in the
unoccupied region (S). Both of these peaks are magnified ten times for clarity
and a broadening of $0.3$ has been used for plotting purpose.
The crossing at $\omega\approx 3$ does not yield any spectral feature
since Im$\Sigma_\mathrm{B}$ at that energy is very large.
}
\label{fig:Sigma_dimer}%
\end{center}
\end{figure}

In terms of the effective potential defined in Eq. (\ref{eq:Xi})
the equations of motion in the bonding-antibonding basis can be recast as
\begin{align}
    \left[i\partial_t-\varepsilon_\mathrm{A}-\Xi_\mathrm{A}(t)
    \right] G_\mathrm{A}(t)
    =\delta(t),
\end{align}
\begin{align}
    \left[i\partial_t-\varepsilon_\mathrm{B}-\Xi_\mathrm{B}(t)
    \right] G_\mathrm{B}(t)
    =\delta(t),
\end{align}
where according to the definition in Eq. (\ref{eq:Xi})
\begin{align}
    \Xi_\mathrm{A}(t)&=V_\mathrm{AA,AA}(t)+V_\mathrm{AB,BA}(t)
    \frac{G_\mathrm{B}(t)}{G_\mathrm{A}(t)},
\\
    \Xi_\mathrm{B}(t)&=V_\mathrm{BB,BB}(t)+V_\mathrm{BA,AB}(t)
    \frac{G_\mathrm{A}(t)}{G_\mathrm{B}(t)}.
\end{align}
These yield
\begin{align}
    \Xi_\mathrm{A}(t>0)    = \frac{\alpha U}{2},
    \qquad
        \Xi_\mathrm{A}(t<0)=-\frac{U}{2\alpha},
\end{align}
\begin{align}
    \Xi_\mathrm{B}(t<0)    = -\frac{\alpha U}{2},
    \qquad
        \Xi_\mathrm{B}(t>0)=\frac{U}{2\alpha}.
\end{align}
Note that the expressions for
$\Xi_\mathrm{A}(t<0)$ and $\Xi_\mathrm{B}(t>0)$ are valid for $U>0$ since
if $U=0$ then $G_\mathrm{A}(t<0)=G_\mathrm{B}(t>0)=0$. 
The effective potential $\Xi$ takes a very simple form, namely
a constant in each time segment and in striking contrast to the
traditional self-energy shown in Fig. \ref{fig:Sigma_dimer}
for the bonding orbital.
The self-energy in the frequency domain
$\Sigma_\mathrm{B}(\omega)$ is defined as
\begin{align}
    G_\mathrm{B}(\omega)=\frac{1}
    {\omega-\varepsilon_\mathrm{B}-\Sigma_\mathrm{B}(\omega)}.
\end{align}
In the dynamical xc potential formalism,
$\Xi_\mathrm{B}(t<0)$ simply shifts the
bonding energy $\varepsilon_\mathrm{B}$ by $-\frac{\alpha U}{2}$
and as a consequence of the many-electron interactions
$\Xi_\mathrm{B}(t>0)$ generates a new feature in the unoccupied part
of the spectrum separated in energy from $\varepsilon_\mathrm{B}$ by
$\frac{U}{2\alpha}$.

\subsection{Homogeneous electron gas}

For the homogeneous electron gas, $V_\mathrm{xc}^\mathrm{0}$ is a constant
and may be set to zero so that
$\Delta V=V^\mathrm{xc}$. 
The natural choice for the orbitals is the plane waves,
\begin{align}
    \varphi_k(r)=\frac{e^{i\mathbf{k}\cdot\mathbf{r}}}{\sqrt{\Omega}}.
\end{align}
The quasiparticle wavefunction
for a given momentum $k$ is given by
\begin{align}
    \psi^*_k(r,t)=G_k(t) 
    \frac{e^{-i\mathbf{k}\cdot\mathbf{r}}}{\sqrt{\Omega}}.
\end{align}
Multiplying Eq. (\ref{eq:EOMQP}) by
$e^{-i\mathbf{q}\cdot\mathbf{(\mathbf{r-r'})}}/\Omega$, 
integrating over $r$ and $r'$, and using
\begin{align}
    \int d^3R\, e^{i\mathbf{q}\cdot\mathbf{R}}=(2\pi)^3\delta(\mathbf{q}),
\end{align}
the equation of motion for $t\neq 0$ and a given spin becomes
\begin{align}
    (i\partial_t - \varepsilon_q)G_q(t)-\frac{1}{\Omega}\sum_\mathbf{k}
    V^\mathrm{xc}_{|\mathbf{q-k}|}(t) G_k(t)=0,
\end{align}
where
\begin{align}
    V^\mathrm{xc}_{|\mathbf{q-k}|}(t)
    &=\frac{1}{\Omega^2}\int d^3r d^3r' 
    e^{-i(\mathbf{q-k})\cdot (\mathbf{r-r'})}
    V^\mathrm{xc}(|\mathbf{r-r'}|,t)
    \nonumber\\
    &=\frac{1}{\Omega} \int d^3R\, e^{-i(\mathbf{q-k})\cdot \mathbf{R}}
    V^\mathrm{xc}(R,t).
\end{align}
The equation of motion can be recast as
\begin{align}
    \left[i\partial_t - \varepsilon_q-\Xi_q(t)\right]G_q(t)=0,
\end{align}
where
\begin{align}
    \Xi_q(t)= \frac{1}{\Omega}\sum_\mathbf{k}
    V^\mathrm{xc}_{|\mathbf{q-k}|}(t) \frac{G_k(t)}{G_q(t)}.
\end{align}
For $t<0$, $\Xi_q(t)$ may be approximated by
\begin{align}
    \Xi_q(t<0)\approx \Xi^\mathrm{S}_q - \Xi^\mathrm{D}_q e^{i\omega_\mathrm{p}t},
\end{align}
where
\begin{align}\label{eq:plasmon}
    \omega_\mathrm{p}=\sqrt{4\pi \rho}
\end{align}
is the plasmon energy of the electron gas with density $\rho$.
A simple model for $\Xi^\mathrm{S}_q$ is
\begin{align}\label{eq:VqS}
    \Xi^\mathrm{S}_q=(1-\gamma Z)(E_\mathrm{F}-\varepsilon_q),
\end{align}
where $Z$ is the quasiparticle renormalization factor which
reduces the bandwidth and $\gamma>1$ is a factor
which takes into account band broadening due to
the momentum-dependence of the self-energy.
Both $Z$ and $\gamma$ are assumed to be momentum independent. 
In this model, the noninteracting occupied bandwidth
is reduced by a combined factor $\gamma Z$, and the new Fermi level
has been realigned with that of the noninteracting one. 

The solution is given by
\begin{align}
    G_q(t)&=G_q(0)e^{-i(\varepsilon_q+\Xi^\mathrm{S}_q)t 
    +i\int_0^t dt' \,\Xi^\mathrm{D}_q e^{i\omega_\mathrm{p}t'}}
    \nonumber\\
    &=G_q(0) e^{-iE_q t}\left( A_0 + A_1e^{i\omega_\mathrm{p}t}
    +A_2 e^{i2\omega_\mathrm{p}t}+...\right),
\end{align}
where
\begin{align}
    E_q=\varepsilon_q+\Xi_q^\mathrm{S}
\end{align}
is the quasiparticle energy and
\begin{align}
    A_0&= 1-\lambda + \frac{1}{2}\lambda^2,\\
    A_1&=\lambda(1-\lambda),\\
    A_2&=\frac{1}{2}\lambda^2,\quad 
    \lambda=\frac{\Xi_q^\mathrm{D}}{\omega_\mathrm{p}}.
\end{align}
Note that $A_0+A_1+A_2=1$.

Assuming that $G_q(0)=1$ 
the coefficient of the dynamical part $\Xi^\mathrm{D}_q$
is then determined by the renormalization of the quasiparticle
$Z=A_0$, yielding
\begin{align}\label{eq:Z}
    \lambda=1-\sqrt{2Z-1}.
\end{align}
When Fourier transformed, $\mathrm{Im}G_q(\omega)$ yields the
expected peaks at the quasiparticle energy $E_q$ and at multiples of
the plasmon satellite energy below the quasiparticle energy, 
$E_q-\omega_\mathrm{p}$, $E_q-2\omega_\mathrm{p}$, ..., which agrees
with the result obtained from the cumulant expansion
\cite{hedin1980,almbladh1983,aryasetiawan1996,kas2014,kas2019}.
In general, $\Xi_q^\mathrm{S}$ is complex, and its imaginary part gives
a life-time broadening of the quasiparticle and the plasmon satellite.
To take into account the lifetime of the quasiparticle,
a model which depends linearly on the inverse lifetime may be employed:
\begin{align}\label{eq:eta}
    \eta(q)=\eta_0 + (\eta_1-\eta_0)\frac{E_\mathrm{F}-\varepsilon_q}
    {E_\mathrm{F}-\varepsilon_0}.
\end{align}
At $q=k_\mathrm{F}$ the broadening is taken to be $\eta_0$ rather than zero
for the purpose of plotting curves.
With $G_q(0)=1$ the momentum-dependent spectrum
is then given by the imaginary part of
\begin{align}\label{eq:ImGq}
G_q(\omega)&=\frac{A_0}{\omega-E_q-i\eta(q)}
+\frac{A_1}{\omega-E_q+\omega_\mathrm{p}-i\eta(q)}
\nonumber\\
&\quad +\frac{A_2}{\omega-E_q+2\omega_\mathrm{p}-i\eta(q)}.
\end{align}
The simple model assumes that for $q\leq k_\mathrm{F}$ the spectrum does not
have weight above the Fermi level. 
\begin{figure}[t]
\begin{center} 
\includegraphics[scale=0.6, viewport=3cm 8cm 17cm 20cm, clip]
{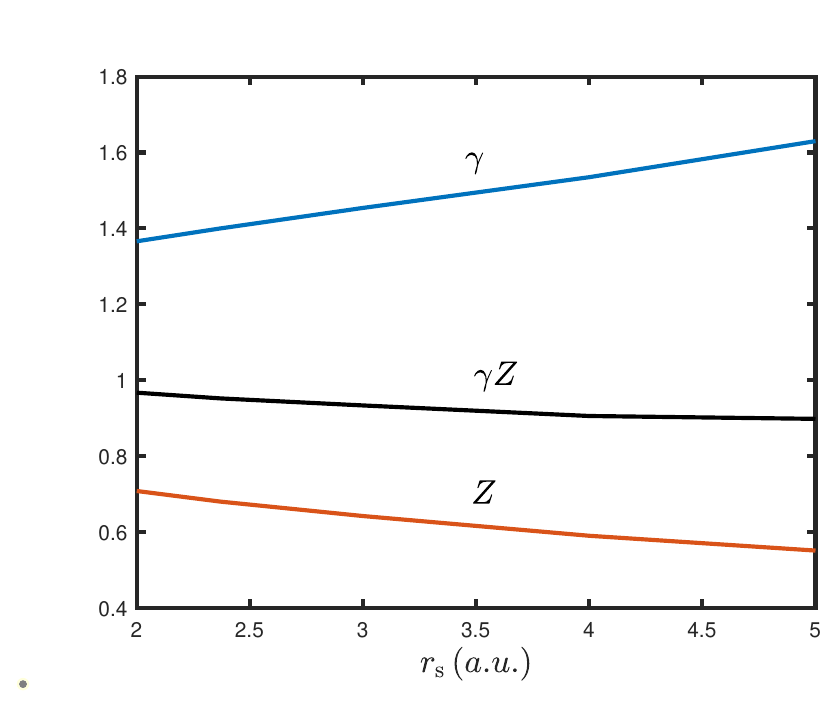}
\caption{The momentum-broadening factor $\gamma$ and
the quasiparticle renormalization factor $Z$ as described in the text
extracted from $GW$ calculations of the homogeneous electron gas.
$\gamma Z$ corresponds to the band narrowing.
The renormalization factor $Z$ is taken to be the average of the values
at $q=0$ and $q=k_\mathrm{F}$.
}
\label{fig:gammaZ}%
\end{center}
\end{figure}
\begin{figure}[t]
\begin{center} 
\includegraphics[scale=0.6, viewport=3cm 8cm 17cm 20cm, clip]
{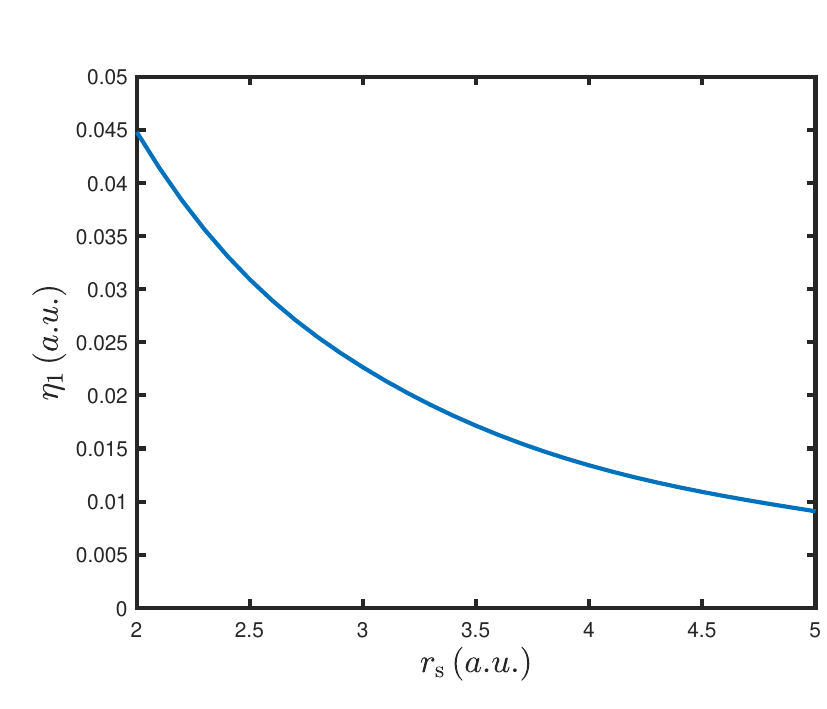}
\caption{The lifetime broadening factor ($\eta_1$ in Eq. (\ref{eq:eta}))
obtained from $GW$ calculations
of the homogeneous electron gas as a function of $r_\mathrm{s}$.
}
\label{fig:eta}%
\end{center}
\end{figure}
The parametrized $\gamma$ and $Z$ as functions of $r_\mathrm{s}$
are shown in Fig. \ref{fig:gammaZ}
and the life-time broadening $\eta_1$ is shown in Fig. \ref{fig:eta}.
They are extracted from one-shot $GW$ \cite{hedin1965} calculations of the 
homogeneous electron gas. The combined factor
$\gamma Z$ provides a measure of band narrowing which is stronger,
the lower the
density (larger $r_\mathrm{s})$, as expected since correlations become
more important.

\section{Applications to Na and Al}

The model given in Eq. (\ref{eq:ImGq}) is now applied to 
solid Na and Al.
The aim is to demonstrate that the proposed model reproduces the
spectra rather well.
Na is bcc with lattice constant $a=4.29$ \r{A} and $1s$ valence electron
per unit cell yielding $r_\mathrm{s}=4.0$ a.u. whereas Al is fcc with
lattice constant $a=4.05$ \r{A} and $2s$ valence electrons per unit cell
yielding $r_\mathrm{s}=2.37$ a.u. 

\begin{figure}[t]
\begin{center} 
\includegraphics[scale=0.6, viewport=3cm 8cm 17cm 20cm, clip]
{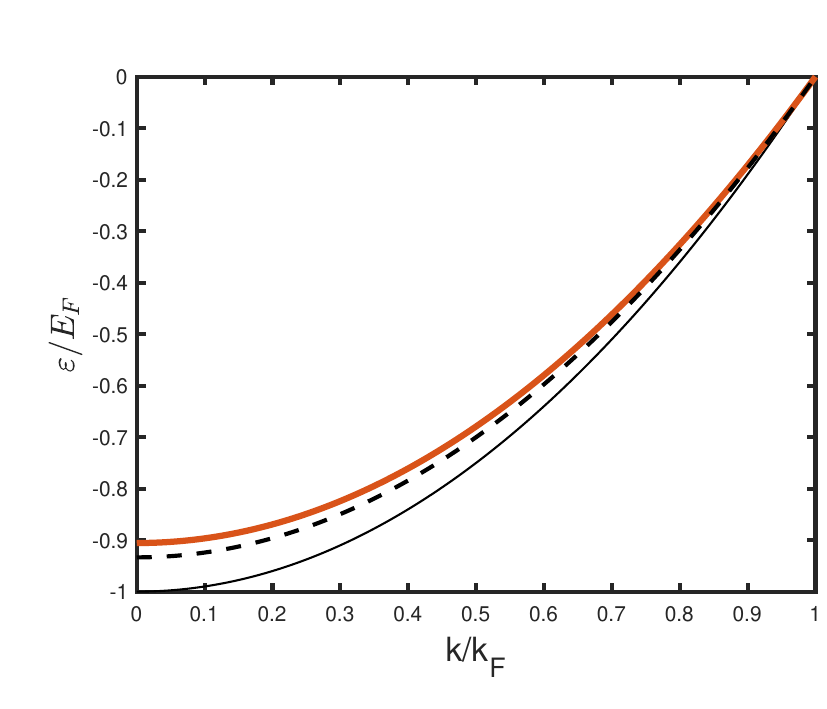}
\caption{The energy dispersion with $r_\mathrm{s}=4.0$ a.u.
corresponding to Na (thick solid line) 
and $r_\mathrm{s}=2.37$ a.u. corresponding to Al (dashed line) 
calculated using the model described in the text
compared with that of the
noninteracting electron gas (thin solid line).
}
\label{fig:dispersion}%
\end{center}
\end{figure}

\begin{figure}[t]
\begin{center} 
\includegraphics[scale=0.6, viewport=5.3cm 1.5cm 20cm 16cm, clip]
{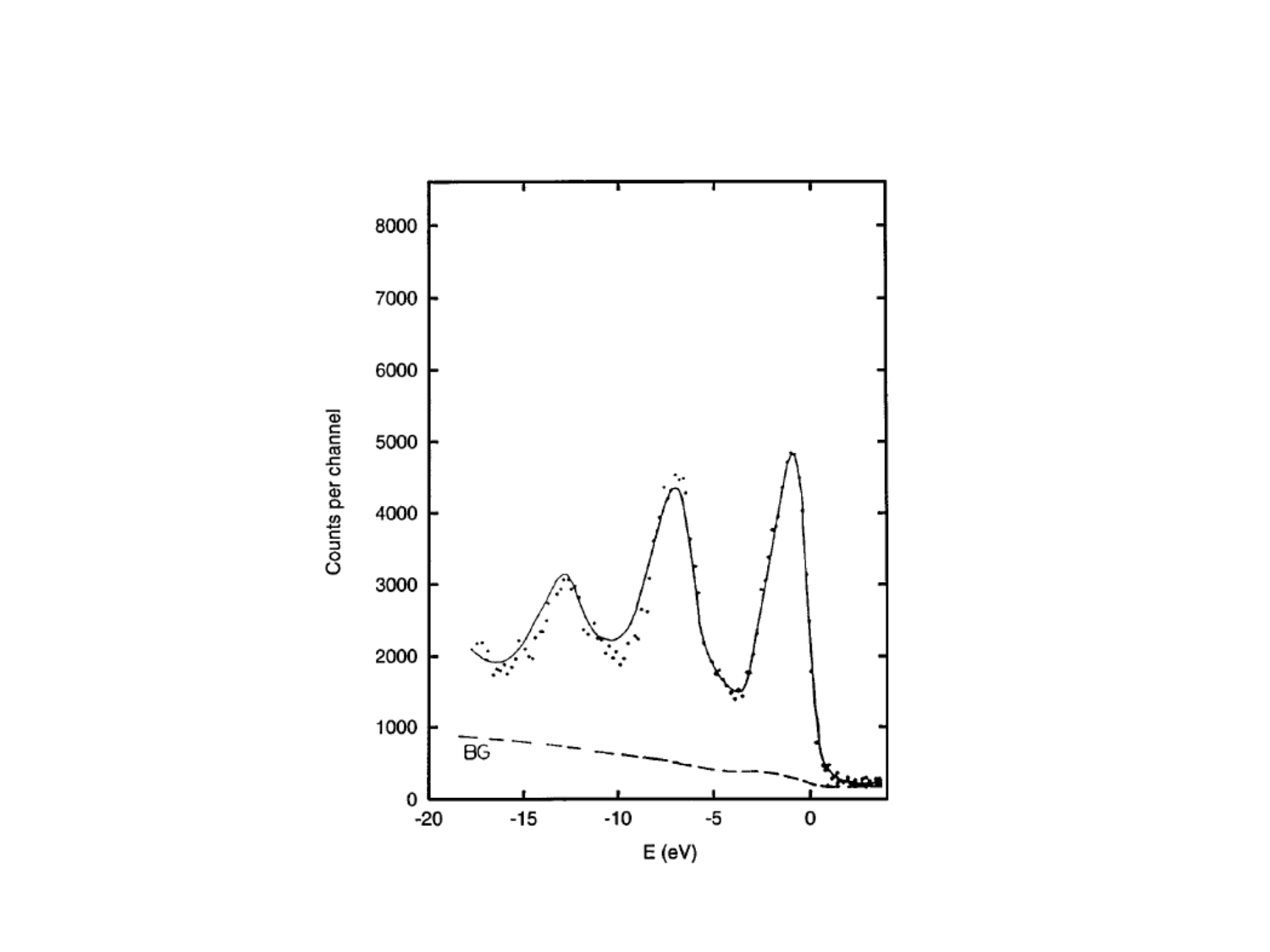}
\includegraphics[scale=0.5, viewport=3cm 8cm 17cm 19cm, clip]
{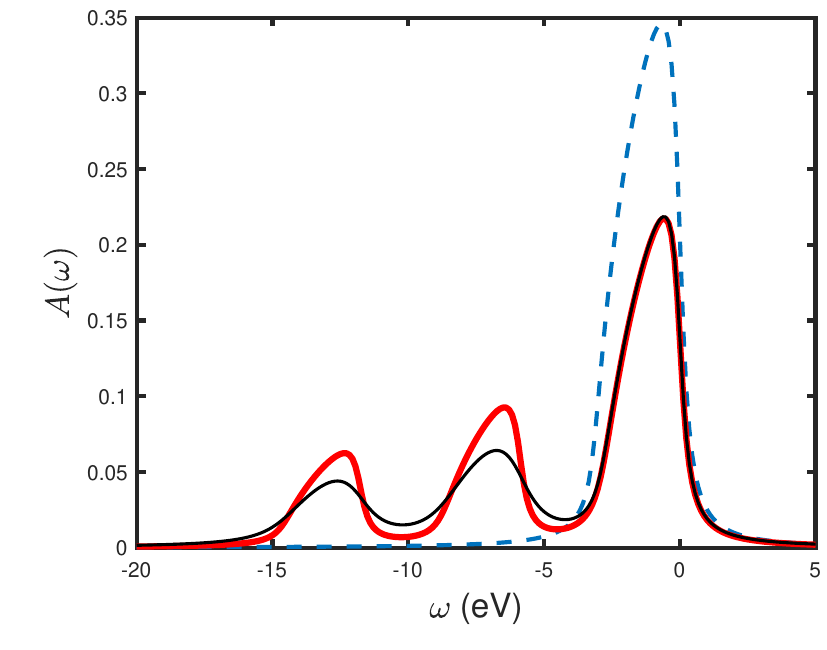}

\caption{The total spectral function of Na: experiment \cite{steiner1979} (top)
and calculated using the model with $r_\mathrm{s}=4.0$ (bottom).
The thin solid line
is the spectrum when a plasmon broadening of $0.55$ eV is included.
For comparison, the noninteracting electron gas total 
spectrum corresponding to $r_\mathrm{s}=4.0$ is also shown (dashed line).
}
\label{fig:AtotNa}%
\end{center}
\end{figure}

\begin{figure}[t]
\begin{center} 
\includegraphics[scale=0.6, viewport=3cm 8cm 17cm 20cm, clip]
{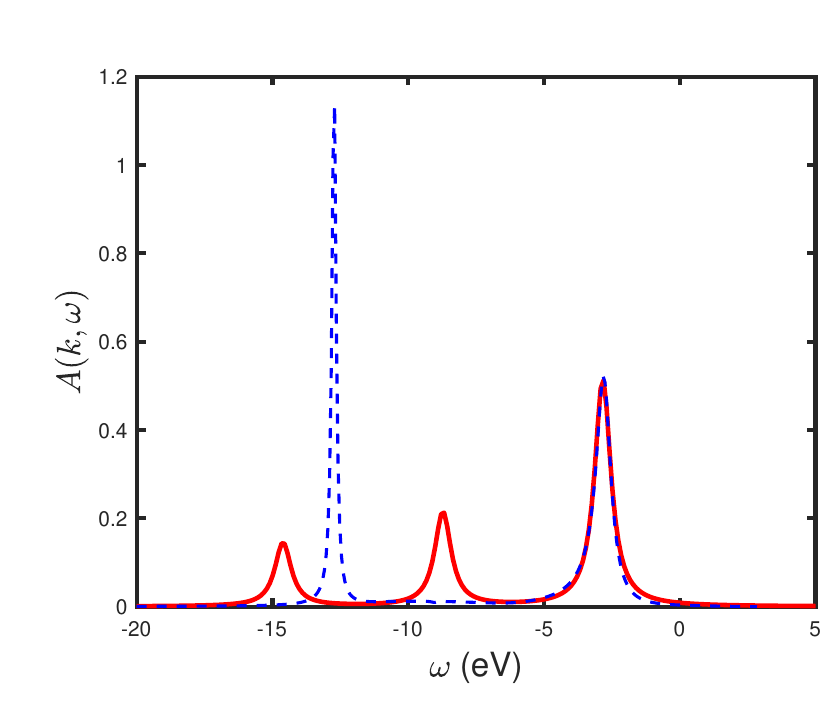}
\caption{The spectral function 
$A(k,\omega)=\frac{1}{\pi}\mathrm{Im}G(k,\omega)$ 
of Na at the $\Gamma$-point ($k=0$) calculated using the model
described in the text (solid line).
The dashed line is the $GW$ spectral function of the homogeneous electron gas
with $r_\mathrm{s}=4.0$.
}
\label{fig:Aq=0Na}%
\end{center}
\end{figure}

\begin{figure}[t]
\begin{center} 
\includegraphics[scale=0.6, viewport=3cm 8cm 17cm 20cm, clip]
{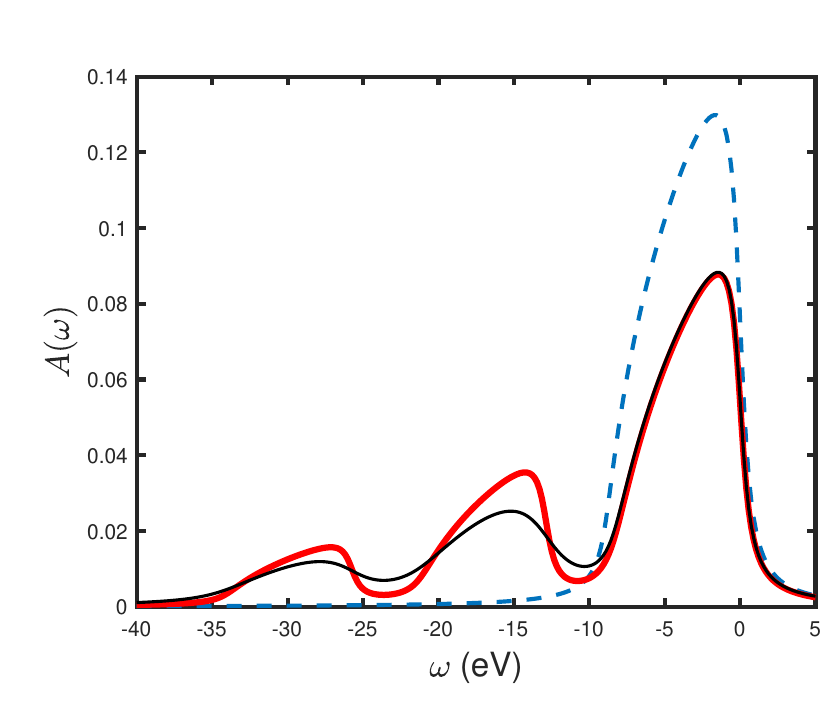}
\caption{The total spectral function of Al calculated using the model
with $r_\mathrm{s}=2.37$ (thick solid line). The thin solid line
is the spectrum when a plasmon broadening of $1.3$ eV is included.
For comparison, the noninteracting electron gas total 
spectrum corresponding to the same $r_\mathrm{s}$ value is shown (dashed line).
}
\label{fig:AtotAl}%
\end{center}
\end{figure}

\begin{figure}[t]
\begin{center} 
\includegraphics[scale=0.6, viewport=3cm 8cm 17cm 20cm, clip]
{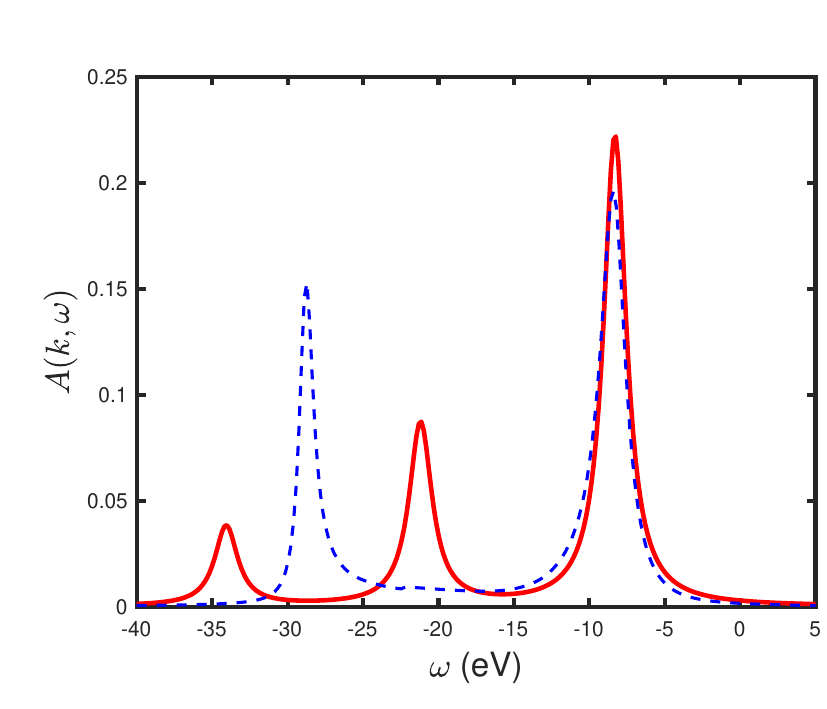}
\caption{The spectral function 
$A(k,\omega)=\frac{1}{\pi}\mathrm{Im}G(k,\omega)$ 
of Al at the $\Gamma$-point ($k=0$) calculated using the model
described in the text (solid line).
The dashed line is the $GW$ spectral function of the homogeneous electron gas
with $r_\mathrm{s}=2.37$.
}
\label{fig:Aq=0Al}%
\end{center}
\end{figure}

The energy dispersions corresponding
to $r_\mathrm{s}=4.0$ (Na) and $r_\mathrm{s}=2.37$ (Al) are
shown in Fig. \ref{fig:dispersion} and compared with the
noninteracting dispersion. The total and $q=0$ spectral functions
are shown in Figs. \ref{fig:AtotNa}
and \ref{fig:Aq=0Na} for Na and
in Figs. \ref{fig:AtotAl} and \ref{fig:Aq=0Al} for Al.

The calculated total spectral function of Na shown in Fig. \ref{fig:AtotNa}
exhibits the characteristic
main quasiparticle peak followed by plasmon satellites separated
by multiples of plasmon energy. It is in close agreement
with the experimental result \cite{steiner1979},
and with the result obtained using the cumulant expansion
method \cite{aryasetiawan1996}. 
To show the effect of plasmon lifetime, an additional broadening
of approximately ten percent of the plasmon energy has been included. This
additional lifetime tends to shift the plasmon peak to a slightly lower energy.

Fig. \ref{fig:Aq=0Na} shows an example of the calculated 
spectral function at the $\Gamma$ point ($q=0$) compared with the result
obtained from the standard one-shot $GW$ calculation.
The $GW$ spectral function suffers from a well-known overestimation
of the plasmon binding energy.

The total and momentum-resolved spectral functions of Al shown in
Figs. \ref{fig:AtotAl} and \ref{fig:Aq=0Al}, respectively,
display qualitatively similar characteristics to those of Na.
The results agree well with those calculated using
the cumulant expansion \cite{aryasetiawan1996}.
The higher average electron density in Al results in the plasmon energy,
which is approximately twice that of Na.


\section{Local-density approximation}

From the ansatz in Eq. (\ref{eq:VqS}), a possible 
local-density approximation \cite{kohn1965,jones1989,becke2014} for metals is
\begin{align}
    \Xi_q^\mathrm{S}(r)= \left[1-\gamma(\overline{\rho})
    Z(\overline{\rho})\right]  
    \left( \frac{1}{2}\left[ 3\pi^2\rho(r)\right]^{2/3}
    -\varepsilon_q\right),
\end{align}
where
\begin{align}
    \overline{\rho}=\frac{1}{\Omega}\int dr \,\rho(r)
\end{align}
is the average density.
Both $\gamma$ and $Z$ can be calculated as functions of the
electron gas density $\overline{\rho}$ within the $GW$ approximation \cite{hedin1965} or
can be treated as fitting parameters. The plasmon energy
$\omega_\mathrm{p}$ is given by
Eq. (\ref{eq:plasmon}) with $\rho=\overline{\rho}$
which together with $Z$ determines $\Xi_q^\mathrm{D}$ using Eq. (\ref{eq:Z}) .

\section{Summary and conclusions}

The proposed definition of the quasiparticle wavefunction offers a natural
description of a quasiparticle decaying in time due to the many-electron
interactions and the coupling
to the collective modes. The matrix elements of the Green function
in a chosen set of one-particle orbitals can be understood as
a coefficient of expansion of the quasiparticle wavefunction in
these orbitals. The quasiparticle wavefunction depends on the
choice of orbitals. In practice, this choice is dictated by
a reference mean-field noninteracting Green function, which in many cases is
given by the Kohn-Sham one.
Irrespective of the orbital choice, it is shown that the quasiparticle 
wavefunction has in general
a total probability distribution less than unity, reflecting the decay.
Indeed, in a noninteracting system, the total probability is unity.

Within the dynamical xc potential formalism,
the quasiparticle wavefunction fulfills a physically appealing equation
of motion in which the dynamics of the quasiparticle is
governed by a time-dependent effective potential. 
In many physical problems, only a subspace of the full Hamiltonian is
mostly relevant. In that case, the quasiparticle wavefunctions can be
solved only for a limited set of quantum numbers $q$. For example,
in a periodic crystal $q$ represents the Bloch momentum $\mathbf{q}$
and the band index $n$, 
and only a limited set of bands relevant to the physics of the problem
may need to be considered.

Examples of the Hubbard dimer
and the electron gas indicate that the effective potential $\Xi_q$ 
has a simple generic form consisting of a time-independent term
and in the case of the electron gas an additional
dynamic term whose time dependence is determined by the
collective excitations. A similar form has also been found
in the one-dimensional Hubbard model \cite{aryasetiawan2022b} 
as well as the one-dimensional Heisenberg spin model \cite{zhao2023}.
The simple form of the effective potential
may be related to the robustness of the orbital representation
\cite{hedin1995}.

A simple ansatz is proposed for metals, in particular the alkali.
Two examples, Na and Al, are given to illustrate the quality
of the proposed model.
The experimental spectrum of Na is well reproduced by the model.
The peak positions and the overall features of Na and Al spectra
calculated using the model agree well with those obtained
from the cumulant expansion method \cite{aryasetiawan1996}.
Naturally, the simple model
cannot capture fine details originating from bandstructure effects,
which are not included in the model. The parameters have been
extracted from $GW$ calculations on the homogeneous electron gas and
improved approximations beyond $GW$ would provide better parameters.

A local-density approximation is also suggested, and its implementation
is the next step in the development of the dynamical xc
potential formalism. 

\begin{acknowledgments}
Financial support from
the Swedish Research Council (Vetenskapsrådet, VR, Grant No. 2021\_04498)
and Carl Tryggers Stiftelse (CTS 23:2419) are gratefully acknowledged.
We thank Rex Godby for valuable discussions.
\end{acknowledgments}

\appendix

\section{Quasiparticle wavefunction in the nonequilibrium case}
\label{app:noneq}

The concept of the quasiparticle wavefunction can be extended to 
the nonequilibrium case in which the Hamiltonian is time dependent.
The time-ordered zero-temperature Green function is defined as
\begin{align}
    iG(rt,r't)=\langle\Psi_0| \mathrm{T}\hat{\psi}(rt) 
    \hat{\psi}^\dagger(r't')|\Psi_0\rangle,
\end{align}
in which $\mathrm{T}$ is the time-ordering symbol, $|\Psi_0\rangle$ 
is some initial state, and $\hat{\psi}(rt)$ is the field operator in
the Heisenberg picture,
\begin{align}
    \hat{\psi}(rt)=\hat{U}(0,t) \hat{\psi}(r) \hat{U}(t,0).
\end{align}
The time-evolution operator $\hat{U}(t,t')$ is given by
\begin{align}
    \hat{U}(t,t')=\mathrm{T} e^{-i\int_{t'}^t dt_1 \,\hat{H}(t_1)},
\end{align}
where $\hat{H}(t)$ is the Hamiltonian of the system which 
can be time dependent.

Choosing a complete set of one-particle orbitals $\{\varphi_k\}$
the Green function can then be written as
\begin{align}
    G(rt,r't) &= \sum_{kk'} \varphi_{k}(r)G_{kk'}(tt')\varphi^*_{k'}(r'),
\end{align}
where
\begin{align}
    iG_{kk'}(tt') =\langle\Psi_0| \mathrm{T}\hat{c}_k(t) 
    \hat{c}^\dagger_{k'}(t')|\Psi_0\rangle.
\end{align}

Defining 
\begin{align}
    \psi^*_k(r',tt')&=\sum_{k'} G_{kk'}(tt')\varphi^*_{k'}(r')
\end{align}
the Green function can be expressed as
\begin{align}
    G(rt,r't) &=  \sum_k \varphi_k(r) \psi^*_k(r',tt').
\end{align}

As in the equilibrium case $\psi^*_k(r',tt')$
can be interpreted as a quasiparticle wavefunction.
One first shows that
\begin{align}
    \sum_{k'} |G_{kk'}(tt')|^2 \leq 1.
\end{align}
Consider the case $t<t'$,
\begin{align}
    |G_{kk}(tt')|^2 &= \left|
    \langle \Psi_0 | \hat{c}_k^\dagger(t')
    \hat{c}_k(t) | \Psi_0\rangle \right|^2
    \nonumber\\
    &= \left|
    \langle \Psi_0 |\hat{U}(0,t') \hat{c}_k^\dagger \hat{U}(t',t)
    \hat{c}_k \hat{U}(t,0)| \Psi_0\rangle
    \right|^2.
\end{align}
Let $|\phi_k (t)\rangle = \hat{c}_k | \Psi(t)\rangle$, where
\begin{align}
    | \Psi(t)\rangle =\hat{U}(t,0)| \Psi_0\rangle.
\end{align}
For fermions
\begin{align}
    \langle \phi_k(t) |\phi_k(t)\rangle 
    = \langle \Psi(t) |\hat{n}_k |\Psi(t)\rangle \leq 1.
\end{align}
Since the time-evolution operator is unitary, it follows that
\begin{align}
    |G_{kk}(tt')|^2 = \left|
    \langle \phi_k(t') |\hat{U}(t',t) | \phi_k(t)\rangle
    \right|^2 \leq 1.
\end{align}
The unitary operator $\hat{U}(t',t)$ rotates
the state $ | \phi_k(t)\rangle$ so that
the magnitude of the overlap of the rotated state with
$\langle \phi_k(t')|$ 
must be less than $\langle \phi_k(t') |\phi_k(t')\rangle$.

For a given $t$ and $t'$, let $S$ be the unitary matrix that diagonalizes $G$:
\begin{align}
    G_{kk'}(tt') = 
    \sum_{k_1} S_{kk_1} \widetilde{G}_{k_1}(tt') S^\dagger_{k_1k'}.
\end{align}
One finds
\begin{align}
   &\sum_{k'} |G_{kk'}(tt')|^2 
   \nonumber\\
   &= \sum_{k'} \sum_{k_1k_2}
   S_{kk_1} \widetilde{G}_{k_1}(tt') S^\dagger_{k_1k'} S_{k'k_2}
   \widetilde{G}^*_{k_2}(tt') S^\dagger_{k_2k}
   \nonumber\\
   &=\sum_{k_1}S_{kk_1} |\widetilde{G}_{k_1}(tt')|^2 S^\dagger_{k_1k}
   \nonumber\\
   &=\sum_{k_1} |\widetilde{G}_{k_1}(tt')|^2 |S_{kk_1}|^2.
\end{align}
Since $\sum_{k_1}|S_{kk_1}|^2=1$ and
$|\widetilde{G}_{k_1}(tt')|^2 \leq 1$ it follows that
\begin{align}
    \sum_{k'} |G_{kk'}(tt')|^2 \leq 1.
\end{align}
As in the equilibrium case, this result shows that
\begin{align}
    \int dr\, |\psi_k(r,tt')|^2 = \sum_{k'} |G_{kk'}(tt')|^2 \leq 1
\end{align}
and permits the interpretation of $\psi^*_k(r,tt')$
as quasiparticle wavefunction. $G_{kk'}(tt')$ can be understood as the
expansion coefficient of the quasiparticle wavefunction
$\psi^*_k(r,tt')$ in the base orbital
$\varphi_{k'}$. A similar analysis can be performed for $t>t'$.


\begin{thebibliography}{99}                                                       

\bibitem {fetter-walecka}See, for example, A. L Fetter and J. D Walecka,
\emph{Quantum Theory of Many-Particle Systems},
(Dover, Mineola, New York, 2003).

\bibitem{lundqvist} See, for example, L. Hedin and S. Lundqvist, 
in \textit{Solid State Physics}, 
edited by H. Ehrenreich, F. Seitz and D. Turnbull
(Academic, New York, 1969), Vol. 23, p. 1.

\bibitem {aryasetiawan2022a}F. Aryasetiawan, 
Phys. Rev. B \textbf{105}, 075106 (2022).

\bibitem {aryasetiawan2022b}F. Aryasetiawan and T. Sj\"ostrand, 
Phys. Rev. B \textbf{106}, 045123 (2022).

\bibitem{karlsson2023} K. Karlsson and F. Aryasetiawan, 
Phys. Rev. B \textbf{107}, 115172 (2023).

\bibitem{aryasetiawan2023} F. Aryasetiawan, 
Phys. Rev. B \textbf{108}, 115110 (2023).

\bibitem{zhao2023} Z. Zhao, C. Verdozzi, and F. Aryasetiawan, 
Phys. Rev. B \textbf{108}, 235132 (2023).

\bibitem {negele}See, for example, J. W. Negele and H. Orland, 
\emph{Quantum Many-Particle Systems}, (Westview Press, Boulder, Colorado, 1998).

\bibitem {dvorak2021} M. Dvorak,
arXiv:2101.00704 [cond-mat.str-el].

\bibitem{slater1951} J. C. Slater, 
Phys. Rev. \textbf{81}, 385 (1951).

\bibitem {slater1968}J. C. Slater, 
Phys. Rev. \textbf{165}, 658 (1968).


\bibitem {kohn1965}W. Kohn and L. J. Sham, 
Phys. Rev. \textbf{140}, A1133 (1965).

\bibitem {jones1989}R. O. Jones and O. Gunnarsson, 
Rev. Mod. Phys. \textbf{61}, 689 (1989).

\bibitem {becke2014}A. D. Becke, 
J. Chem. Phys. \textbf{140}, 18A301 (2014).

\bibitem {hedin1980}L. Hedin, 
Phys. Scr. \textbf{21}, 477 (1980).

\bibitem{almbladh1983} C.-O. Almbladh and L. Hedin, 
in \emph{Handbook on Synchroton Radiation},
edited by E. E. Koch (North-Holland, Amsterdam, 1983) Vol. 1, p.686.

\bibitem{aryasetiawan1996} F. Aryasetiawan, L. Hedin, and K. Karlsson,
Phys. Rev. Lett. \textbf{77}, 2268 (1996).


\bibitem{kas2014} J. J. Kas, J. J. Rehr, and L. Reining,
Phys. Rev. B \textbf{90}, 085112 (2014).

\bibitem{kas2019} J. J. Kas, T. D. Blanton, and J. J. Rehr,
Phys. Rev. B \textbf{100}, 195144 (2019).

\bibitem {hedin1965}L. Hedin, 
Phys. Rev. \textbf{139}, A796 (1965).

\bibitem{steiner1979}P. Steiner, H. Höchst, and S. Hüfner, 
in \emph{Photoemission in Solids II}, edited by L. Ley and M. Cardona, 
Topics in Applied Physics Vol. 27 (Springer-Verlag, Heidelberg,1979).

\bibitem{hedin1995} L. Hedin, 
Int. J. Quantum Chem. \textbf{56}, 445 (1995).


































\end{thebibliography}
\end{document}